\documentclass[11pt,showkeys]{revtex4-2}
\usepackage{relsize}
\usepackage{hyperref}
\usepackage{amsmath,amsfonts,amssymb}
\hypersetup{dvips,dvipdfm,colorlinks=true,urlcolor=magenta,filecolor=magenta,linktoc=page,citecolor=red,linkcolor=blue,bookmarks=true}
\usepackage{graphicx,epstopdf}
\usepackage[latin1]{inputenc}
\usepackage{tikz}
\usetikzlibrary{shapes,arrows}
\usepackage{bm}
\usepackage{hyperref}
\usepackage{amsmath,amsfonts,amssymb}
\hypersetup{dvips,dvipdfm,colorlinks=true,urlcolor=magenta,filecolor=magenta,linktoc=page,citecolor=red,linkcolor=blue,bookmarks=true}
\usepackage{graphicx,epstopdf}
\usepackage{array}
\usepackage{hyperref}

\usepackage{hyperref}
\usepackage{graphicx}
\usepackage{dcolumn}
\usepackage{color}
\usepackage{bm}
\hypersetup{dvips,dvipdfm,colorlinks=true,urlcolor=magenta,filecolor=magenta,linktoc=page,citecolor=red,linkcolor=blue,bookmarks=true}
\usepackage{amsmath}
\usepackage{amsfonts}
\usepackage{amssymb}
\usepackage{epstopdf}
\usepackage{float}

\input epsf

\begin{document}

\title{ Non-equilibrium thermodynamics in  the non-canonical scalar field perturbed space-time : Stability analysis}

\author{Subhayan Maity\footnote {maitysubhayan@gmail.com}}
\affiliation{Department of Mathematics, Jadavpur University, Kolkata-700032, West Bengal, India.}


\begin{abstract}
	The space-time of the Universe has been perturbed under  a scalar field $\phi$
 considering the minimum coupling  between  $\phi$ and the background metric. The solutions of Einstein field equations have been obtained under perturbed geometry and the corresponding conservation equation shows the non-equilibrium thermodynamic prescription of the cosmic fluid.  Following the stability criteria of the cosmic fluid along with the laws of thermodynamics, some constraints have been imposed on the choice of $\phi$.  

\end{abstract}
\keywords{Particle creation, Non-canonical scalar field}

\maketitle




\section{Introduction} \label{sec1}
The popular explanation for the recent  cosmic evolution pattern is the introduction of cosmological constant $\lambda$.  Although it fits excellently with the observation, it suffers from two serious issues namely coincidence problem \cite{Steinhardt:2003st} and cosmological constant problem \cite{Weinberg:2000yb,Padmanabhan:2002ji} from theoretical point of view. Now it is claimed that the vacuum energy associated with $\lambda$ is not constant, rather it decays into the other  matter components . In some models, $\lambda $ is replaced by  some phenomenological choices for variable cosmological constant or cosmological scalar field $\phi$  to resolve the above the issues. In some models, the interaction between $\phi$ and the cosmic fluid has been interpreted as a non-equilibrium thermodynamic process with a dissipative cosmic fluid. Under such non-equilibrium thermodynamic prescription, various transport mechanism like particle creation, diffusion etc.\cite{Benisty:2017eqh,Benisty:2018oyy,Calogero:2011re,Calogero:2012kd,Franchi:10485F,Haba:2009by,Herrmann:024026,Perez:2020cwa,new} have been assumed to set up the theoretical interpretation of the cosmic evolution pattern. In most of the models, the effect of scalar field $\phi$ has been included in the matter component of the Einstein field equation ($R_{\mu \nu}-\frac{1}{2} R g_{\mu \nu} +\phi g_{\mu \nu}=K T_{\mu \nu}$  \cite{emg,con},  the dimension factor of $\phi$ has been chosen as unity for convenience ) with out affecting the background space-time. In some cases,the Klein-Gordon equation of the scalar field with canonical Lagrangian[$\mathcal{L}=\frac{1}{2}\partial  ^{\mu} \phi  \partial _{\mu} \phi-V(\phi)$] has been modified under curved space-time and also coupling with gravitational field\cite{Pimentel:2021ldr} [$(\Box +m^2+\xi R)\phi=0$] .  In this work, it is aimed to exhibit the non-equilibrium thermodynamic phenomena from the perturbation of original metric. The perturbation of space-time is expected form the coupling with matter source. Here the matter source has been chosen in the form of  a non-canonical scalar field.  The coupling with non-canonical scalar field  leads to the perturbation in the background metric\cite{7}. Asper literature, there is possibly no model which describes the perturbation in space-time with canonical scalar field. In this work also,  no change  has been found in the  original  the metric due to coupling with a canonical field ($L= X-V(\phi)$, $X$ is the canonical kinetic term). Hence it is reasonable to choose non-canonical field  because any source with canonical lagrangian can  atbest add an extra source term with the energy-momentum tensor but it can not chnage the background metric.  \par The Friedmann equations under the perturbed space-time yields the conservation equation(non-conservation)  of the cosmic fluid under the non-equilibrium thermodynamic phenomena like particle creation. This work is such a model of perturbed space-time which leads to the non-equilibrium thermodynamic evolution of the Universe .  Also considering the $2$ nd law of thermodynamics along with stability criteria of the cosmic fluid, some restrictions on  $\phi$ have been made in this scenario.

\section{non-equilibrium thermodynamics of cosmic fluid in scalar field perturbed space-time} \label{sec2}
Non-canonical scalar field is a general k-essence field with non-canonical kinetic term like $L=T(X,\phi)-V(\phi)$ with $T\neq X$  where $X =\frac{1}{2} g^{\mu \nu} \nabla _ \mu \phi \nabla _\nu \phi$, the canonical kinetic term.

The non-canonical  scalar field $\phi$ minimally coupled to the background metric $g_{\mu \nu}$ yields the action \cite{1,2,3,4,5,6,7},
\begin{equation}
S_k [\phi ,g_{\mu \nu}]=\int d^4 x\sqrt{-g} L(X,\phi).     \label{1}
\end{equation}  Here the energy momentum tensor is given by,
  \begin{equation}
  T_{\mu \nu}= - L_X \nabla _ \mu \phi \nabla _\nu \phi+ g_{\mu \nu} L \label{2}
  \end{equation} with $L_X=\frac{\partial L}{\partial X}$. Now applying variational principle, one can find the equation of motion $( -\frac{1}{\sqrt{-g}}\frac{\delta S_k}{\delta \phi} =0)$ in the form,
  \begin{equation}
 G^{\mu \nu} \nabla _ \mu \nabla _\nu \phi + 2 X \frac{\partial^2 L}{\partial X \partial \phi}-\frac{\partial L}{\partial \phi} =0,   \label{3}
  \end{equation} where $G^{\mu \nu}= \frac{1}{{L_X}^2 \left (1+2\frac{\partial \ln L_X}{\partial \ln X}\right)^{\frac{1}{2}}}[L_X g^{\mu \nu}+L_{XX}\nabla ^{ \mu }\phi \nabla ^{\nu} \phi]$  with $L_{XX}=\frac{\partial ^2 L}{\partial X^2}$ and obviously $\left (1+2\frac{\partial \ln L_X}{\partial \ln X}\right)> 0$ for a suitable and well defined Lagrangian. Hence, the inverse metric is given by,
  \begin{equation}
  G_{\mu \nu} =L_X \left (1+2\frac{\partial \ln L_x}{\partial \ln X}\right)^{\frac{1}{2}}\left [g_{\mu \nu} - \frac{L_{XX}}{L_x + 2 X L_{XX}} \nabla_\mu \phi \nabla_\nu \phi \right ]= \frac{L_X}{C_s}\left [g_{\mu \nu} - \frac{L_{XX}}{L_x + 2 X L_{XX}} \nabla_\mu \phi \nabla_\nu \phi \right ],   \label{4}
  \end{equation} where $C_s = \left (1+2\frac{\partial \ln L_X}{\partial \ln X}\right)^{-\frac{1}{2}}$, the sound speed. After a conformal  transformation \cite{a,b}, one gets the metric of perturbed space-time as
  \begin{equation}
  \tilde{G_{\mu \nu}}= \frac{C_s}{L_X} G_{\mu \nu}=g_{\mu \nu} - \frac{L_{XX}}{L_X + 2 X L_{XX}} \nabla_\mu \phi \nabla_\nu \phi  \label{5} .
  \end{equation} 

In general a non-canonical scalar field  has  a general DBI-type \cite{8} of Lagrangian $L(X,\phi) = 1-f(\phi)\sqrt{1-2X}$. Gangopadhyay, Manna et al. in several works\cite{a,b,c,d}, used such lagrangian for successful exhibition of emergent gravity and other phenomena. Some modifications are also found in general DBI lagrangian for incorporating  the effect of K-essence.  
 
  In this work, the nonlinear Dirac -Born- Infield   type of Lagrangian has been chosen in the form $L=- m(\phi)\sqrt{1-2X}-v(\phi)$ . This form of lagrangian corresponds to the relativistic particle in Minkwoski space-time. For example $L=-m\sqrt{1-2X}$ for free particle, $L=-m\sqrt{1-2X}-\frac{1}{2}m {\omega}^2 {\phi}^2$  for harmonic oscillator etc.  Hence, one has the simplest form of perturbed metric as,
  \begin{equation}
  \tilde{G_{\mu \nu}}= g_{\mu \nu}-\partial _ \mu  \phi \partial _ \nu \phi   \label{6}
  \end{equation} 

Notably, for canonical lagrangian $L=X-v(\phi)$, one has $\tilde{G_{\mu \nu}}= g_{\mu \nu}$ (no change in original metric)  .

  \par In this work, the background metric is chosen as a flat FLRW Universe  [$ds^2=g_{\mu \nu}dx^{\mu} dx^{\nu}=dt^2- a^2(t)(dx^2+dy^2+dz^2)$ ,with $a(t)$ as the scale factor of the Universe ] and hence the perturbed metric with minimally coupled scalar field can be found in the form $(d\tilde{s}^2=\tilde{G_{\mu \nu}}dx^{\mu} dx^{\nu})$ of equation (\ref{6}),
  \begin{equation}
  d\tilde{s}^2=(1-\dot{\phi}^2)dt^2 -\left[(a^2+\phi _x ^ 2)dx^2+(a^2+\phi _y ^ 2)dy^2+(a^2+\phi _z ^ 2)dz^2 \right]-2 \left[\dot{\phi} \phi _xdxdt + \dot{\phi} \phi _ydydt+\dot{\phi} \phi _zdzdt \right ] ,  \label{7} 
  \end{equation} where $\dot{\phi}=\frac{\partial \phi}{\partial t}$ and $\phi _{x_i}=\frac{\partial \phi}{\partial x_i}$. If the the scalar field $\phi$ is chosen to be the explicit function of time only i.e. $\phi =\phi(t)$, then the perturbed metric can be simplified as,
  \begin{equation}
  d\tilde{s}^2=(1-\dot{\phi}^2)dt^2-a^2(dx^2+dy^2+dz^2)   \label{8}
  \end{equation}
   Here for a well defined space-time, $\dot{\phi}^2<1$ must be followed.
   \par  One may apply Einstein field equation $R_{\mu \nu}-\frac{1}{2}R g_{\mu \nu}=8 \pi G T_{\mu \nu}$( taking speed of light in vacuum $c=1$) to obtain the evolution of the thermodynamic parameters of the cosmic fluid under perturbed space-time
      with  $R_{\mu \nu}$ as Ricci tensor and $R=\tilde{G^{\mu \nu}} R_{\mu \nu} $ being the Ricci scalar, $G$ is the gravitational constant and can be taken as $1$ for suitable scaling of unit. The energy - momentum tensor, $T_{\mu \nu}=\tilde{G_{\alpha \nu}}T^{\alpha}_\mu$ is chosen as a perfect fluid and hence, $T^{\alpha}_\beta=diag(\rho, -P,-P,-P)$. $\rho$ and $P$ are the density and pressure of the cosmic fluid respectively. Hence Einstein field equation yields the modified form of Friedmann equations,
      \begin{equation}
      3 H^2=8 \pi \rho (1-\dot{\phi} ^2)   \label{9}
      \end{equation}
       \begin{equation}
       2 \dot{H}+2H\frac{\dot{\phi}\ddot{\phi}}{1-\dot{\phi}^2}=-8\pi(P+\rho)(1-\dot{\phi}^2),   \label{10}
       \end{equation} with $H=\frac{\dot{a}}{a}$, Hubble parameter and $\dot{H}=\frac{d H}{d t}$. The above modified friedmann equations (\ref{9}) and (\ref{10}) can also be written as (in the form of original Friedmann equations)
       \begin{equation}
       3 H^2=8 \pi \tilde{\rho}~~ \mbox{and} -2\dot{H}=8 \pi (\tilde{P}+\tilde{\rho}+\mathcal{P}), \label{11}
       \end{equation}
     where $\tilde{P}=P(1-\dot{\phi}^2) , ~\tilde{\rho}= \rho (1-\dot{\phi}^2)$ are the effective pressure and density respectively of the cosmic fluid in modified back ground. $\mathcal{P}=\frac{1}{4 \pi} \frac{\dot{\phi} \ddot{\phi}}{1-\dot{\phi}^2}H$ can be treated as a dissipative pressure term arisen due to the effect of perturbation with scalar-field $\phi$. Hence the conservation equation of the modified thermodynamic state of the fluid can be found as,
     \begin{equation}
     \dot{\tilde{\rho}}+3 H(\tilde{\rho}+\tilde{P}+\mathcal{P})=0 ,  \label{12}
     \end{equation} where interestingly the original fluid sustains its conservation equation with out any dissipation $\dot{\rho}+3H(P+\rho)=0$ . 
  \par So, one can claim that under unperturbed space-time the cosmic fluid is a conservative fluid. But due to perturbation in space-time some extra degrees of freedom will be added with the original fluid and the effective fluid system becomes dissipative. Hence under perturbed Universe, even a conservative fluid may execute non-equilibrium thermodynamic phenomena (example- particle creation - annihilation process)  under suitable conditions.
  \par Effectively, such a system can be treated as a two fluid system with a non-dissipative barotropic fluid ( with density $\rho$) and a dissipative barotropic fluid ( with density $\rho^{(\phi)}=-\rho \dot{\phi}^2$) of same barotropic index $\omega$ (say). Hence, $\frac{P}{\rho}=\frac{P^{(\phi)}}{\rho ^{(\phi)}}=\frac{\tilde{P}}{\tilde{\rho}}=\omega$ and also the effective density of the combination of the two fluids is additive i.e.  $\tilde{\rho}=\rho +\rho^{(\phi)}$. $\rho ^{(\phi )}$ is the energy density of the extra added degrees of freedom induced by the perturbation. The two fluids satisfy the individual conservation relation as, 
  
  \begin{subequations}
  	\begin{equation}
  	\dot{\rho}+3(1+\omega)H \rho =0  ~~~~~ \mbox{and}~~~     \label{13a} 
  	 \end{equation}  
  	 \begin{equation}
  	\frac{\partial \rho ^{(\phi)}}{\partial  t}+\left [3(1+\omega)H -2\frac{\ddot{\phi}}{\dot{\phi}}\right]\rho ^{(\phi)} =0.     \label{13b}
  	\end{equation}
  \end{subequations} 
Now equations (\ref{13a}) and (\ref{13b}) along with equation (\ref{11}) yield 
\begin{equation}
2\dot{a}\ddot{a}+(3\gamma -2)\frac{\dot{a}^3}{a}+2  \rho _0 a^{(2-3\omega)} \dot{\phi} \ddot{\phi} =0,  \label{14}
\end{equation} where $\gamma=(1+\omega)$ and $\rho _0 = \rho ({t=t_0})$, $t_0$ being a reference epoch of time. The equation (\ref{14}) can also be written in terms of Hubble parameter as,
\begin{equation}
2\dot{H}+3\gamma H^2+ 2 H \frac{\dot{\phi}\ddot{\phi}}{1-\dot{\phi}^2}=0   \label{15}
\end{equation}
 
\section{Particle creation mechanism}

The non-equilibrium thermodynamics of cosmic fluids can be described as particle creation-annihilation process . Lima introduced \cite{9}an unknown non-collisional term $\mathcal{P}_g$ as an extra source term in Boltzman equation to incorporate the effect of gravitation induced particle creation process.
\begin{equation}
	p^{\mu} \frac{\partial f}{\partial x ^{\mu}}-\Gamma ^{\mu}_{\alpha \beta}p^{\alpha }p^{\beta} \frac{\partial f}{\partial p^{\mu}} =C(f)+\mathcal{P}_g,  \label{16}
\end{equation}
 
 where $\Gamma ^{\mu}_{\alpha \beta}$ are christoffel symbols and $C(f)$ is the total collisional term. The above boltzman equation (\ref{16}) describes the evolution of phase space density $f(x^{\mu}, p^{\mu})$(here $\mu =0,1,2,3$). The particle number density is related to $f$ as 
 \begin{equation}
 	n=\frac{g}{(2\pi)^3}\int f(x^{\mu},p^{\mu}) d^3 p ,  \label{17}
 \end{equation} 
with $g$, the number of spin degrees of freedom. Under mass shell condition, $f$ can be taken independent of temporal momentum $(p^0)$ i.e. $f=f(x^{\mu}, p^{i})$ with $i=1,2,3$.
Hence the equation (\ref{16}) can be assumed in simplified form under unperturbed space-time as, 
\begin{equation}
	\bar{p}^0 \frac{\partial f}{\partial x^0} - H\bar{p}^0 \bar{p}^i \frac{\partial f}{\partial \bar{p}^i}=C(f)+\mathcal{P}_g \label{18}.
\end{equation} Here $\bar{p}^0=E$ and $\bar{p} ^i =ap^i$ .
  Following Lima's intuition \cite{9}$\mathcal{P}_g =-\lambda \frac{\Gamma }{3H}\Gamma ^{\mu}_{\alpha \beta}p^{\alpha }p^{\beta} \frac{\partial f}{\partial p^i}$, one gets 
  \begin{equation}
  	\frac{\partial f}{\partial t}=3H\left (1-\frac{\Gamma }{3H} \right) p\frac{\partial f}{\partial p}, \label{19}
  \end{equation} after neglecting the collisional term $C(f)$. $\Gamma  $ is the particle creation rate.  Now  following equation (\ref{17}), one has the particle number non-conservation equation
\begin{equation}
	\dot{n}+3H(1-\frac{\Gamma }{3H})n=0  \label{20}
\end{equation}

\par In this work, an attempt is taken to incorporate the particle creation process in the boltzman equation without introducing any extra source term. Instead of taking any intuition , Boltzman equation has been simplified under perturbed geometry (equation (\ref{8})) and one finds 
\begin{equation}
	E\left [\frac{\partial f}{\partial t}-3H\sqrt{1-\dot{\phi}^2}p\frac{\partial f}{\partial p} \right ]=C(f) .  \label{21}
\end{equation}
Now imposing equation (\ref{17}), one finds 
\begin{equation}
	\frac{d n}{d t}+3H\sqrt{1-\dot{\phi}^2}n =\frac{g}{(2\pi)^3}\int(f) \frac{d^3 p}{E}. \label{22}
\end{equation}

Here also neglecting the effect of collisional term, the particle number non-conservation equation can be written as,
\begin{equation}
	\frac{d n}{d t}+3H\sqrt{1-\dot{\phi}^2}n =0  .  \label{23}
\end{equation}

Clearly the effect of perturbation leads to the dissipation in particle number conservation with particle creation rate $\Gamma _\phi = 3H \left (1-\sqrt{1-\dot{\phi}^2}\right )$. Also comparing with Lima's intuition, one may relate the scalar field with the extra particle creation source term  in perturbed space-time as 
\begin{equation}
	\left (\mathcal{P}_g \right )_{\phi}= -\lambda \left (1-\sqrt{1-\dot{\phi}^2}\right )\Gamma ^{\mu}_{\alpha \beta}p^{\alpha }p^{\beta} \frac{\partial f}{\partial p^i}  \label{23a}
\end{equation}

Hence the gravitation induced particle creation process can be assumed as a result of perturbation of underlying space-time by a scalar field. Hence this phenomena can be termed as scalar field induced particle creation mechanism.
\section{Emergence of cosmic space and entropy correction, 2nd law of thermodynamics.}
The effect of the scalar field on the entropy of the system (Universe) will be studied through the Padmanabhan's equation regarding the emergence of cosmic space. In the present context, the energy density is taken additive hence the degrees of freedom will also be additive following the equipartition law of energy. Therefore the original equation for an expanding system \cite{10,11} $\left[\frac{d \mathcal V}{dt}\propto \left(N_{\mbox{Surrounding}}-N_{\mbox{Bulk}}\right )\right]$
is modified for a two fluid system as,
\begin{equation}
\frac{d \mathcal V}{dt}= \left [\left(N_{\mbox{Surrounding}}+N_{\mbox{Surrounding}} ^{(\phi)}\right ) -\left(N_{\mbox{Bulk}}+N_{\mbox{Bulk}} ^{(\phi)}\right ) \right ]   \label{24}
\end{equation}, taking the proportionality constant as unity. Here the degrees of freedom on the surface of the Hubble sphere, $N_{\mbox{Surrounding}}= 4 S$, $S$ being the entropy on the horizon and $N_{\mbox{Bulk}}= \frac{2 |E_{\mbox{Komar}}|}{T}$, is the degrees of freedom in the bulk of the unperturbed system. The terms $N_{\mbox{Surrounding}} ^{(\phi)}$ and $N_{\mbox{Bulk}} ^{(\phi)}$ are the extra degrees of freedom on the surface and in the bulk respectively arisen due to the scalar field. \par Now, the komar energy of the original fluid, $E_{\mbox{Komar}}=-\frac{4}{3} \pi (3 P+\rho)\frac{1}{H^3} $ , the horizon temperature, $T=\frac{2 \pi}{H}$ and hence one may find from the equations (\ref{9}) and (\ref{24}),
\begin{equation}
-\frac{\ddot{a}}{a}=\frac{H^4}{4 \pi}\left[N_{\mbox{surrounding}}-N_{\mbox{Bulk}}\right] +\frac{4}{3} \pi (3P+\rho).   \label{25}
\end{equation}
 For the effective dissipative fluid, one can also take, 
 \begin{equation}
 N_{\mbox{Bulk}}^{(\phi)}=2\frac{| E_{\mbox{Komar}}^{(\phi)}|}{T}=-\frac{16 \pi^2 }{3 H^4}\left(3 P^{(\phi)}+\rho ^{(\phi)}\right).   \label{26}
 \end{equation}
 Hence one obtains the form of $N_{\mbox{Surrounding}}^{(\phi)}$ from equations (\ref{10}), (\ref{25}) and (\ref{26}) as,
 \begin{equation}
 N_{\mbox{Surrounding}}^{(\phi)}=\frac{4 \pi}{H^3}.\frac{\dot{\phi} \ddot{\phi}}{1-\dot{\phi}^2} .   \label{27}
 \end{equation} 
 Now the effective entropy  of the system is found as,
 \begin{equation}
 S_{\mbox{eff}}=\frac{N_{\mbox{Surrounding}}+N_{\mbox{Surrounding}}^{(\phi)}}{4}=S\left (1+\frac{\dot{\phi} \ddot{\phi}}{(1-\dot{\phi}^2)H}\right),   \label{28 }
 \end{equation}
 where $S=\frac{\pi}{H^2}$, is the entropy of the unperturbed Universe.
 \par In the present context, the scalar field Lagrangian is assumed like a relativistic free particle as
 \begin{equation}
 	L=-m\sqrt{1-\dot{\phi}^2}      \label{29}
 \end{equation}
 where $m$ is the mass of the particle. Hence from equation (\ref{3}), one finds for the Lagrangian of equation (\ref{29})
 \begin{equation}
 	3 H+\frac{\ddot{\phi}}{\dot{\phi}(1-\dot{\phi}^2)}=0.   \label{30}
 \end{equation}
Therefore the effective entropy of the Universe in such condition will be,
\begin{equation}
	 S_{\mbox{eff}}=S(1-3\dot{\phi}^2)  \label{31}
\end{equation}
 
  To satisfy the $2$nd law of thermodynamics $\left(\dot{S}_{\mbox{eff}}>0 \right)$, one has
  \begin{equation}
  	\frac{\dot{S}}{S}- \frac{6\dot{\phi} \ddot{\phi}}{1-3\dot{\phi}^2} >0   \label{32}
  \end{equation}
The above equation (\ref{32}) can also be written asper equation (\ref{30}) as
\begin{equation}
	9H\frac{\dot{\phi}^2(1-\dot{\phi}^2)}{1-3 \dot{\phi}^2}-\frac{\dot{H}}{H} >0   \label{33}
\end{equation}
Now from equation(\ref{15}) and (\ref{30}), equation (\ref{33}) is found as 
\begin{equation}
	\dot{\phi}^2-3\frac{\dot{\phi}^2(1-\dot{\phi}^2)}{1-3 \dot{\phi}^2}< \frac{\gamma}{2}  \label{34}
\end{equation}
Hence one may constrain the value of $\dot{\phi}$ as
\begin{equation}
	F<\frac{1+\omega}{3 \omega  +7} ,   \label{35}
\end{equation} where  $F=\dot{\phi} ^2$.
\section{THERMODYNAMIC ASPECTS OF SCALAR FIELD : STABILITY ANALYSIS.}
The thermodynamic features of a cosmic fluid depend on the sign of 
 the thermodynamic derivatives like specific heats ($C_p, C_v$), compressibilities ($K_s, K_T$) and expansibility $(\alpha)$ of that fluid. Within an isolated system which undergoes thermodynamic work done due to the continuous expansion of volume, a stable fluid must possess the positive values of all these thermodynamic derivatives i.e. \cite{12,13}
 \begin{equation}
 	C_p , C_v,K_T, K_S, \alpha \geq 0 .   \label{36}
 \end{equation}
\subsection{SPECIFIC HEATS OF COSMIC FLUID.}
The specific heats of a fluid can be obtained from the relations
\begin{equation}
	C_v=\left(\frac{\partial E}{\partial T} \right )_v ~,~ C_p=\left( \frac{\partial h}{\partial T}\right)_p     \label{37}
\end{equation} 
where $E=\tilde{\rho} v$, total internal energy of the fluid and $h=E+\tilde{P}V= (1+\omega) E$, enthalpy of that fluid. $T$ is the absolute temperature of the universe. From equation(\ref{13a}), one has the solution of energy density of the conservative fluid as, 
\begin{equation}
	\rho =\rho _0 a^{-3(1+\omega)}  ,     \label{38}
\end{equation} with $\rho _0 =\rho (t_0)$. $t_0 $ is a reference epoch of time with $a(t_0)=1$.  Hence the expression of $\tilde{\rho}$ can be found as
\begin{equation}
	\tilde{\rho}=(1-\dot{\phi}^2)\rho _0 a^{-3(1+\omega)}     \label{39}
\end{equation}
The above equation (\ref{39})  can be written in the differential form as
\begin{equation}
	d \ln \tilde{\rho}=-(1+\omega)d \ln v + d \ln (1-\dot{\phi}^2)    \label{40}
\end{equation} where $v=v_0 a^3$, is the volume of the universe with $v_0=v(t=t_0)$, volume of the universe at reference epoch.
Again asper first law of thermodynamics and idea of entropy, one has
\begin{equation}
	dS=\frac{v}{T}\frac{d \tilde{\rho}}{d T} dT +(1+\omega )\frac{\tilde{\rho}}{T} d v   \label{41}
\end{equation}
Considering entropy $S$ as state variable of volume ($v$) and temperature ($T$), one finds
\begin{equation}
	d \ln T =\frac{1+\omega}{\omega} d \ln \tilde{\rho}.   \label{42}
\end{equation} Therefore, it is found from equations (\ref{40}) and (\ref{42}), 
\begin{equation}
	 d \ln T=-\omega d \ln v + \frac{\omega}{1+\omega}d \ln (1-\dot{\phi}^2)  \label{43}
\end{equation} Again from equations (\ref{40}) and (\ref{43}), one obtains the evolution of energy with temperature as,

\begin{equation}
 E=E_0 \frac{T}{T_0} \left (1-\dot{\phi}^2 \right )^{\frac{1}{1+\omega}} ,     \label{44}	
\end{equation}
with $E_0=E(t_0)$ and $ T_0= T(t_0)$.
Now, One has the time evolution of temperature of universe under particle creation mechanism as,
\begin{equation}
	\frac{\dot{T}}{T}+\omega(3H-\Gamma)=0 .  \label{45}
\end{equation} Hence in the present context, one finds from equations (\ref{23}) and (\ref{45}),
\begin{equation}
	\dot{T}=-3\omega H \sqrt{1-\dot{\phi}^2} ~T . \label{46}
\end{equation} 

\begin{figure}
	\centering
	\includegraphics*[width=0.6\linewidth]{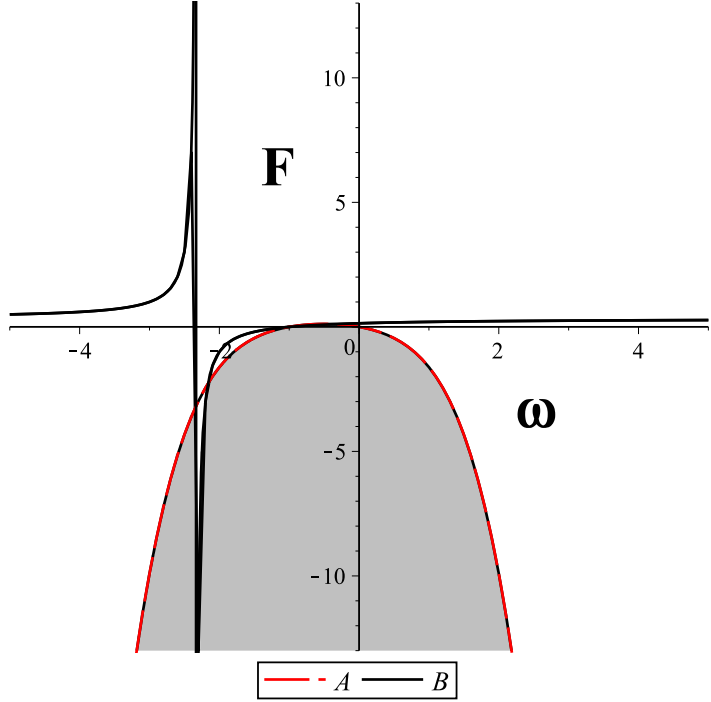}
	
	\caption{Stability region for $\phi,F=\dot{\phi}^2$,A :$\frac{F}{\sqrt{1-F}} =\frac{\omega (1+\omega)}{2}$, B: $	F=\frac{1+\omega}{3 \omega  +7}$, The shaded area for $\omega >0$ represents the stability region for $\phi$. }
\end{figure}

Therefore one can estimate the specific heats at constant volume $(C_v)$ and at constant pressure $(C_P)$ of the fluid in the form from equations (\ref{37}), (\ref{44}) and (\ref{46}) as,
\begin{equation}
	C_v=\frac{E}{T} ~~,~~ C_P=C_v \left[(1+\omega)+\frac{2 \dot{\phi}\ddot{\phi}}{3H\omega (1-\dot{\phi}^2)^{\frac{3}{2}}} +\frac{d \omega}{d \ln T}      \right]  .  \label{47}
\end{equation}
From equation(\ref{30}) $C_P$ can be simplified for relativistic free particle as,
\begin{equation}
	C_P=C_v \left[(1+\omega) -\frac{2 \dot{\phi}^2}{\omega \sqrt{1-\dot{\phi}^2}} +\frac{d \omega}{d \ln T}    \right ]   \label{48}
\end{equation}
which eventually yields following the condition of equation (\ref{43}) 
\begin{equation}
	C_P=C_v \left[(1+\omega) -\frac{2 \dot{\phi}^2}{\omega \sqrt{1-\dot{\phi}^2}} +\frac{d \ln \omega}{-3 d \ln a +\frac{1}{1+\omega}d \ln (1-\dot{\phi}^2)}    \right ]   \label{49}
\end{equation}
\subsection{COMPRESSIBILITY AND EXPANSIBILITY OF THE COSMIC FLUID}
Considering the volume$(v)$ of a system as a function of temperature $(T)$and pressure$(P)$, one has
\begin{equation}
	dv=v\alpha dT- vK_T dP  ,  \label{50}
\end{equation}
	with $\alpha= \frac{1}{v} \left(\frac{\partial v}{\partial T}\right )_P$, the expansibility and $K_T= -\frac{1}{v}\left (\frac{\partial P}{\partial v}   \right)_T$, the isothermal compressibility. So, one may write 
	\begin{equation}
		\frac{\alpha}{K_T}=\left(\frac{\partial P}{\partial T}    \right)_v    \label{51}
	\end{equation}
Following $Pv=\omega C_v T$, one has 
\begin{equation}
	\alpha =\frac{1}{\omega T}\left (\omega+\frac{d \ln \omega}{d \ln T} \right ) =\frac{1}{\omega T} \left (\omega+\frac{d \ln \omega}{-3d \ln a+\frac{1}{1+\omega} d \ln (1-\dot{\phi}^2)} \right ).  \label{52}
\end{equation}
From equation (\ref{51}), it can be easily found that 
\begin{equation}
	K_T=\frac{\alpha v}{\omega C_v}.    \label{53}
\end{equation}
Note that there is a well known thermodynamic relation between specific heats and compressibilities as
\begin{equation}
	\frac{K_S}{K_T}=\frac{C_v}{C_P}.   \label{54}
\end{equation}
Hence, one also has from equations (\ref{53}) and (\ref{54}),
\begin{equation}
	K_S=\frac{\alpha v}{\omega C_P}    \label{55}
\end{equation}
\subsection{CONSTRAINTS ON SCALAR FIELD FROM STABILITY ANALYSIS}
For the cosmic fluids with constant barotropic index$(\omega)$, the expressions of $\alpha$ and $C_p$ can be simplified as,
\begin{equation}
	\alpha =\frac{1}{T}  ~~,~~ C_P=C_v\left[(1+\omega) -\frac{2 \dot{\phi}^2}{\omega \sqrt{1-\dot{\phi}^2}} \right ] .  \label{56}
\end{equation}
 Clearly following the stability conditions of equation (\ref{36}), one can impose the constraint on $\phi$ as,
 \begin{equation}
 	\frac{F}{\sqrt{1-F}} \leq\frac{\omega (1+\omega)}{2}   \label{57}
 \end{equation}

and also $\omega >0$ .
  The condition of equation (\ref{35})  according to the $2$nd law of thermodynamics must  also be followed by scalar field in the stable scenario. Hence equations (\ref{35}) and (\ref{57}) constraints the stability  region of the scalar field. These two conditions are represented graphically and the stability region has been displayed in Fig-1. It is clear that the stable region belongs the negative values of $\dot{\phi}^2$.

  \section{Discussion}
  This work accomplishes a successful attempt  to correlate the effect of non-canonical scalar field and the non-equilibrium thermodynamic  mechanism. The minimal coupling of DBI type scalar field lagrangian with back ground space-time yields the perturbation in the background metric. The solution of Einstein field equation under perturbed metric corresponds to the non-conservation equation of a dissipative cosmic fluid.  The thermodynamic aspects associated with the evolution of perturbed space-time also has been analysed and some restrictions have been imposed on $\phi$ for thermodynamically well behaved  cosmic evolution pattern. 
  The space-time perturbed with real scalar field never be thermodynamically stable in this scenario. Only complex scalar field following the restrictions may be the suitable candidate for thermodynamically stable Universe. Also the dark energy ($\omega <0$) with constant equation of state will not be thermodynamically stable in perturbed universe and this result agrees with previous works on stability criteria of cosmic fluid \cite{12,13}.

  finally, it is hoped that the phenomenological choices for  complex $\phi$ (obeying the restrictions) may have the possible answers  
  to the recent observational data of cosmic evolution scenario.
  
  \section*{Acknowledgements}  The author SM acknowledges UGC for
  awarding Research fellowship and also thanks Prof.Subenoy Chakraborty, Dept. of Mathematics, J.U. for his valuable suggestions.

  \section*{Data availability}   It is the author's intention to make the data that have been used to determine this result and publish this article as free and easily available as possible to permit readers to reproduce these results.       
  

\section*{References}
  \begin{thebibliography}{50}
  	
  	
  	
  		\bibitem{Steinhardt:2003st} 
  	P. J. Steinhardt, 2003,`
  	Phil. Trans. Roy. Soc. Lond. A
  	361, 2497.\\
  	
  	\bibitem{Weinberg:2000yb} S. Weinberg, 1989, 
  	Rev. Mod. Phys.  61, 1.\\
  	
  	
  	
  	\bibitem{Padmanabhan:2002ji} T. Padmanabhan, 2003,  
  	Phys. Rept.  380, 235 
  	
  	\bibitem{Perez:2020cwa}
  	A.~Perez, D.~Sudarsky , E.~Wilson-Ewing, 2021,
  	Gen. Rel. Grav., 53 ,
  	no.1, 7 
  	
  	
  	
  	
  	
  	\bibitem{Benisty:2017eqh}
  	D.~Benisty , E.~I.~Guendelman, 2017,
  	Eur. Phys. J. C ,77, no.6, 396 
  	
  	
  	\bibitem{Benisty:2018oyy}
  	D.~Benisty, E.~Guendelman and Z.~Haba, 2019, 
  	Phys. Rev. D 99, no.12, 123521 
  	\bibitem{Franchi:10485F}
  	J.Franchi, Y.Le Jan., 2017 ,
  	Comm. Pure Appl. Math.,60:187-251 .
  	
  	\bibitem{Herrmann:024026}
  	J.Hermann., 2010,  Phys. Rev. D, 82:024026 .
  	
  		\bibitem{new}
  	
  	Maity Subhayan
  	Chakraborty Subenoy;
  	2021 , 
  	International Journal of Modern Physics A .
  	
  	\bibitem{Calogero:2011re}
  	S.~Calogero, 2011, 
  	JCAP 11, 016 
  	
  	\bibitem{Calogero:2012kd}
  	S.~Calogero, 2012, 
  	J. Geom. Phys. 62, 2208-2213 
  	\bibitem{Haba:2009by}
  	Z.~Haba, 2010, 
  	Class. Quant. Grav. 27, 095021 
  	
  		
  		
  	\bibitem{emg}	
  	S.Maity, S.Chakraborty, 2022, I.J.M.P.A.37, no. 3
  		
  		\bibitem{con}	
  	S.Maity, S.Chakraborty, 2021, I.J.M.P.A.36, no. 29	
  	\bibitem{Pimentel:2021ldr}
  	L.~O.~Pimentel ,  F.~Pineda, 2021
  	Gen. Rel. Grav. 53, no.7, 62 
  	
  	\bibitem{7}
  	Alexander Vikman, 2007 ,
  	K-essence: Cosmology, causality
  	and Emergent Geometry, Dissertation an der Fakultat
  	fur Physik,Arnold Sommerfeld Center for Theoretical
  	Physics, der Ludwig-Maximilians-Universitat Munchen,
  	Munchen .
  	
  		\bibitem{4}
  	E.Babichev, V.Mukhanov and A.Vikman, 2006, JHEP 09, 061
  	
  	\bibitem{5}
  	E.Babichev,M.Mukhanov and A.Vikman, 2008,  JHEP 0802
  	101 .
  	
  	
  	
  	
  	\bibitem{6}
  	
  	E.Babichev,M.Mukhanov \&  A.Vikman, WSPCProceedings, February 1, 2008.
  	
  	 \bibitem{3}
  	L.P.Chimento, 2004, Phys.Rev.D69 123517 .
  	
  	
  		\bibitem{1}
  	M.Visser,C.Barcelo , S.Liberati, 2002,  Gen.Rel.Grav. 34
  	1719 .
  	
  		\bibitem{2}
  	R.J. Scherrer, 2004,  Phys.Rev.Lett.93 011301 .
  	
  	
  	\bibitem{a}
  	D.Gangopadhyay, G.Manna , Eur.Phys. Lett.100,49001(2012)
  	
  	\bibitem{b}
   G.Manna ,	D.Gangopadhyay, Eur.Phys. Lett.74,2811(2014)
  		\bibitem{8}
  	M.Born and L.Infeld,Proc.Roy.Soc.Lond, 1934,  A144
  	425.
  	
  	\bibitem{c}
  	G.Manna, Eur.Phys. Lett.80,9(2020)
  	\bibitem{d}
  	G.Manna , B.Mazumder, A.Das, Eur.Phys. Lett.135,1(2020)
  	\bibitem{9}
  	J.~A.~S.~Lima and I.~Baranov, 2014, 
  	Phys. Rev. D 90, no.4, 043515 .
  	
  	\bibitem{11}
  	
  	E.~Chang-Young, 2016, 
  	PoS FFP14, 077 .
  	
  	\bibitem{10}
  	K.~Yang, Y.~X.~Liu and Y.~Q.~Wang, 2012, 
  	Phys. Rev. D 86, 104013 .
  	
  	
  	
  	
  
  	
  	
  	
  	
  	\bibitem{12}
  	E.~M.~Barboza, R.~C.~Nunes, E.~M.~C.~Abreu \& J.~A.~Neto, 2015, 
  	Phys. Rev. D 92, no.8, 083526 .
  	
  		\bibitem{13} 
  	S.~Maity, P.~Bhandari , S.~Chakraborty, 2019, 
  	Eur.\ Phys.\ J.\ C  79, no. 1, 82 
  	
  	
  	
  	
  	
  	
  	
  	
  	
  	
  	
  	
  	
  	
  	
  	
  	
  	
  	
  	
  	
  	
  	
  	
  	
  	
  	
  	
  	
  	
  	
  	
  	
  	
  	
  	
  	 
  	
  	
  	
  
  	
  	
  	
  	
  	
  	
  	
  
  	
  
  
  	
  	
  	
  	
  	
  	
  	
  	
  	
  	
  	
  	
  	
  	
  	
  	
  	
  	
  	    
  	
  	
  	
  	
  	
  	
  
  	
  	
  	
  	
  	
  	
  	
  	
  		
  
  	
  	
  
  	
  	
  	
  	
  	
  	
  	
  	
  	
  
  	
  	
  	
  
  	
  	
  	
  	
  	
  	 
  	 
  
  	 
  	
  	 
  	  
  \end{thebibliography}
\end{document}